\documentclass[12pt]{article}
\usepackage{amscd}
\usepackage{verbatim}
\usepackage{hyperref}
\usepackage{amssymb}
\usepackage{amsmath}
\usepackage{tabularx}
\usepackage{graphicx}
\usepackage{mathtools}
\DeclarePairedDelimiter\bra{\langle}{\rvert}
\DeclarePairedDelimiter\ket{\lvert}{\rangle}
\DeclarePairedDelimiterX\braket[2]{\langle}{\rangle}{#1 \delimsize\vert #2}
\newcommand{\RomanNumeralCaps}[1]
    {\MakeUppercase{\romannumeral #1}}

\graphicspath{ {/User/sam/Documents/Latex/Second law 9 may} }
\begin{document}
\begin{center}
\vspace{24pt} { \large \bf Generalized Entropy in Higher Curvature Gravity And Entropy of Algebra of Observables} \\
\vspace{30pt}
\vspace{30pt}
\vspace{30pt}
{\bf Mohd Ali \footnote{mohd.ali@students.iiserpune.ac.in}}, {\bf Vardarajan
Suneeta\footnote{suneeta@iiserpune.ac.in}}\\
\vspace{24pt} 
{\em  The Indian Institute of Science Education and Research (IISER),\\
Pune, India - 411008.}
\end{center}
\date{\today}
\bigskip
\begin{center}
{\bf Abstract}
\end{center}

Recently, Chandrasekaran, Penington and Witten (CPW) have shown that the generalized entropy of the Schwarzschild black hole at the bifurcation surface equals the entropy of an extended von Neumann algebra of quantum observables in the black hole exterior, in semiclassical Einstein gravity. They also derive a version of the Generalized Second law. We generalize these results to a static black hole in an arbitrary diffeomorphism invariant theory of gravity. Thus, a version of the Generalized second law for an arbitrary diffeomorphism invariant theory of gravity follows.
\newpage
\section{Introduction}
Generalized entropy in Einstein gravity was introduced by Bekenstein in order that the second law of thermodynamics be valid near black holes \cite{JB}, \cite{JB2}. He suggested the Generalized Second Law (GSL) holds, namely that the generalized entropy increases under future evolution along the black hole horizon. The generalized entropy for a quantum black hole coupled to matter, in the semiclassical $G \to 0$ limit was defined to be
\begin{equation}\label{In1}
S_{gen} = <\frac{A}{4\hbar G}> + S_{QFT},
\end{equation}
where $A$ is the black hole horizon area and $S_{QFT}$ is the entanglement entropy of the quantum fields in the black hole exterior. As has been pointed out, both the terms in (\ref{In1}) are individually ultraviolet (UV) divergent (the first term due to loop effects which renormalize $G$ and the second term, entanglement entropy, which is UV divergent), but the sum is UV finite \cite{SU}---\cite{gesteau}. The GSL for Einstein gravity was proved by Wall \cite{AW}.
In an arbitrary diffeomorphism invariant theory of gravity, we can analogously define the generalized entropy for a black hole with quantum fields to be the sum of its horizon entropy $S$ in that theory and the entanglement entropy of quantum fields in the black hole exterior,
\begin{equation}
S_{gen} = S + S_{QFT}.
\end{equation}
A candidate for the horizon entropy for a stationary black hole is Wald entropy \cite{Wald}, \cite{VR}. The Wald entropy is ambiguous for a non-stationary black hole --- these ambiguities were first discussed in \cite{JKM}. A linearized GSL (ignoring gravitons) was proved for Lovelock gravity in \cite{SW}.

In an interesting recent development, the generalized entropy in Einstein gravity appeared as the entropy of a von Neumann algebra of observables in the black hole exterior where the black hole is the Schwarzschild solution in Einstein gravity \cite{VGE}. Leutheusser and Liu \cite{LL} (see also \cite{LL1} ) studied the holographic boundary operator algebra of the CFT dual to gravity in the asymptotically anti-de Sitter (AdS) black hole spacetime. They found a emergent Type  \RomanNumeralCaps {3}$_1$ von Neumann algebra for single trace operators in the large $N$ limit of the boundary CFT (see also \cite{lashkari}). Later, Witten \cite{EW3} showed that by enlarging the set of operators to include the boundary Hamiltonian and enlarging the Hilbert space to include a degree of freedom corresponding to a boundary time shift, the algebra becomes a Type \RomanNumeralCaps {2}$_\infty$ von Neumann algebra. In \cite{VRGE}, Chandrasekaran, Longo, Penington and Witten discuss how this construction can be generalized to asymptotically flat black holes. Starting with quantum fields in the exterior of a Schwarzschild black hole, by including the ADM Hamiltonian in the set of operators and the time shift degree of freedom in the Hilbert space, the algebra of these operators in the bulk spacetime is Type \RomanNumeralCaps {2}$_\infty$. A Type \RomanNumeralCaps {2} von Neumann algebra (unlike a Type  \RomanNumeralCaps {3} algebra) has a notion of a (renormalized) trace, density matrix and corresponding von Neumann entropy associated with the density matrix (for a review of von Neumann algebras and their classification, see \cite{EW1}). In \cite{VGE}, Chandrasekaran, Penington and Witten (CPW) showed that the entropy of semiclassical states in the boundary algebra is equal to the generalized entropy of the black hole at the bifurcation surface. This has a bulk interpretation --- in semiclassical gravity in the $G \to 0$ limit, the generalized entropy is the entropy of the algebra of operators in the black hole exterior. Finally, CPW use the monotonicity of the entropy of the algebra under trace preserving inclusions to prove a version of the Generalized Second Law (GSL) showing monotonicity of generalized entropy between early and late times.

In this paper, we study the generalization of these results to an arbitrary diffeomorphism invariant theory of gravity by also including gravitons. We first write the  black hole entropy in such a theory at an arbitrary horizon cut, which is the Wald entropy \cite{Wald}, \cite{VR} with an extra term representing an ambiguity in the Wald entropy for a non-stationary black hole \cite{JKM}. We work in semiclassical gravity. We consider a static (therefore stationary) black hole that is slightly perturbed due to infalling quantum matter and gravitons. In the limit when the cut $v \to \infty$, the perturbed black hole approaches a stationary black hole ($v$ is the affine parameter along the null generator of the horizon). We compute the entropy at $v \to \infty$ minus the entropy at the bifurcation surface up to quadratic order in the perturbation, and we take into account the contribution due to gravitons to the stress-energy tensor. It is possible to fix the ambiguity in the Wald entropy in such a way that this difference of entropies takes a simplified form proportional to the $vv$ component of the stress-energy tensor. Generalizing the computations of CPW \cite{VGE}, we find the difference of \emph{generalized} entropies at $v \to \infty $ and at the bifurcation surface to be proportional to the relative entropy, which is non-negative. By computing the entropy of the extended von Neumann algebra of the black hole exterior \cite{EW3}, we show that the entropy of the algebra is the generalized entropy at the bifurcation surface just as in \cite{VGE} for Einstein gravity. All the above constructions can be done for asymptotically flat static black holes \cite{VRGE}. Finally, we discuss the monotonicity result of CPW \cite{VGE} who show that monotonicity of relative entropy under trace preserving inclusions can be used to argue that the generalized entropy at late times is more than that at early times. For this, we need asymptotically AdS black holes with a holographic dual, but modulo this change, the monotonicity result of CPW goes through for the generalized entropy of a black hole in an arbitrary diffeomorphism invariant theory of gravity.

In section II, we discuss the difference of entropies at $v \to \infty$ and at the bifurcation surface for a slightly perturbed black hole. We use boost arguments which we summarize in section II.1 to simplify this difference of entropies. By expanding the Raychaudhuri equation order by order in the perturbation parameter, we compute the change in entropies to quadratic order in section II.4 both without graviton contributions to the stress-energy tensor, and with the graviton contribution included. In section III, we discuss the entropy of the algebra of operators in the black hole exterior. We first summarize salient results from earlier papers of Witten \cite{EW3}, Chandrasekaran, Longo, Penington and Witten \cite{VRGE} and CPW \cite{VGE} who discussed how the entropy of the algebra was related to the generalized entropy in Einstein gravity. We then generalize these results to an arbitrary diffeomorphism invariant theory of gravity. We conclude in section IV with a discussion.
\section {Entropy Change in Higher Curvature Theory}
\begin{figure}[h]
  \centering
  \includegraphics[width=0.80\textwidth]{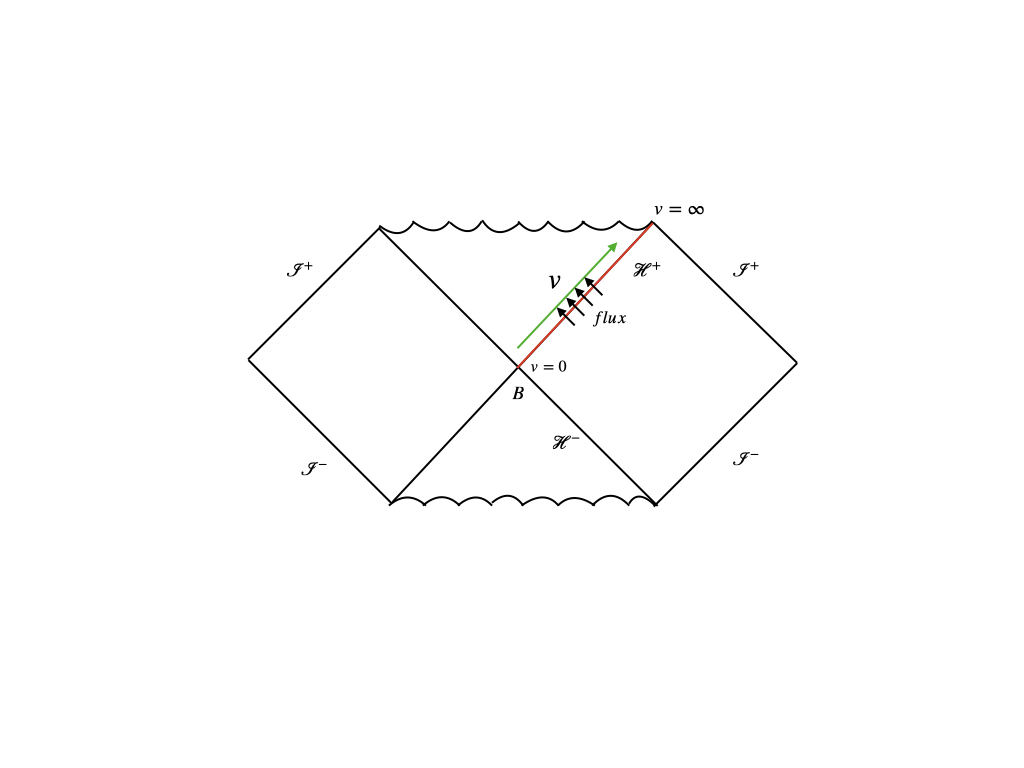}
  \caption{Accretion of matter across the horizon in an asymptotically flat black hole.}
  \label{fig:BH1}
\end{figure}
\vspace{5mm}
In what follows and the rest of the paper, we work in units where $G=1$.
Consider an entropy function for a black hole horizon in an arbitrary diffeomorphism invariant theory of gravity with matter,
\begin{equation}\label{I1}
S= \frac{1}{4}\int \rho \sqrt{h} d^{D-2}x
\end{equation}
where  $\rho = 1 +\rho_w +\Omega$, where $h$ is the induced metric on the  $D-2$ dimensional transverse cut on the horizon and $1 + \rho_w$ is the Wald local entropy density \cite{Wald}, \cite{VR}. As is well-known, the Wald entropy is unambiguously defined for a stationary black hole, but suffers from ambiguities when evaluated on a non-stationary black hole. These ambiguities were pointed out by Jacobson, Kang and Myers (JKM) \cite{JKM} and by Iyer and Wald \cite{VR}. $\Omega$ is a correction to Wald entropy density representing this JKM ambiguity, such that it vanishes for a stationary solution. We are interested in a black hole spacetime with a regular bifurcation surface $\mathcal{B}$, which is slightly perturbed from stationarity by throwing some quantum matter. Let $v$ be an affine parameter along the null generator of the future horizon, such that $v=0$ is the bifurcation surface as shown in Figure \ref{fig:BH1}. Then, the entropy at an arbitrary horizon cut (given by $v= constant$) is
\begin{equation}\label{I2}
S[v]=\frac{1}{4}\int_v \rho \sqrt{h} d^{D-2}x
\end{equation}
where the subscript $v$ in the integral indicates that the integral is over the transverse space at fixed $v$ on the horizon.
We can compute the change in the entropy along the horizon,
\begin{equation}\label{I3}
\frac{dS}{dv}=\frac{1}{4}\int_v \sqrt{h} d^{D-2}x \Big(\frac{d\rho}{dv} +\theta \rho\Big)
\end{equation}
where expansion $\theta \equiv \frac{1}{\sqrt{h}}\frac{d\sqrt{h}}{dv}$. To compute change in the entropy from $v=0$ to $v \rightarrow \infty$, we can integrate both the sides with respect to $v$. This yields
\begin{equation}\label{I4}
\Delta S=  \frac{1}{4} \int_{0}^{\infty} dv \int_v \sqrt{h} d^{D-2}x \Big(\frac{d\rho}{dv} +\theta \rho\Big).
\end{equation}
Here, $\Delta S= S(\infty)-S(0)$. Using integration by parts,
\begin{equation}\label{I5}
\Delta S = \frac{1}{4}\int_v \Big \{ v \sqrt{h}  \Big(\frac{d\rho}{dv} +\theta \rho\Big)\Big\} \Big|_{v=0}^{v \rightarrow \infty} d^{D-2}x \hspace{2mm} - \frac{1}{4} \int_{0}^{\infty} dv \int \sqrt{h} d^{D-2}x v \Big \{\frac{d^2 \rho}{dv^2} + \frac{d\theta}{dv}\rho + 2 \frac{d \rho}{dv}\theta +\theta ^2 \rho \Big \}.
\end{equation}
We will assume that $\sqrt{h}  \Big(\frac{d\rho}{dv} +\theta \rho\Big)$ goes to zero faster than $\frac{1}{v}$, therefore the first term in the above equation is identically zero and we are left with
\begin{equation}\label{I6}
\Delta S = - \frac{1}{4} \int_{0}^{\infty} dv \int \sqrt{h} d^{D-2}x v \Big \{\frac{d^2 \rho}{dv^2} + \frac{d\theta}{dv}\rho + 2 \frac{d \rho}{dv}\theta +\theta ^2 \rho \Big \}.
\end{equation}
To compute $\Delta S $ order by order, we will now consider the metric perturbation sourced by a stress-energy tensor of order $\epsilon$, i.e,  $<T_{vv}> \hspace{2mm}\sim O(\epsilon)$. We will also assume that the perturbation is about a stationary black hole background and at late times, the black hole will again settle down into a stationary state. The perturbation expansion we are interested in is
\begin{equation}\label{I7}
g_{\mu \nu}= g^{(0)}_{\mu \nu} + \epsilon^{\frac{1}{2}} g^{(\frac{1}{2})}_{\mu \nu}+
\epsilon g^{(1)}_{\mu \nu}+ O(\epsilon^{\frac{3}{2}}),
\end{equation}
where the zeroth order term corresponds to the stationary black hole solution with regular bifurcation surface, the $\sqrt{\epsilon}$ term is due to quantized graviton fluctuations, and the $\epsilon$ term is due to the
gravitational field of matter or gravitons. We can think of $\epsilon$ as $\hbar$. We want to emphasize that  $\Omega$  vanishes at order $\sqrt{\epsilon}$ at the bifurcation surface \cite{AS}, a fact which will be useful later in the calculations.
\subsection{Boost Argument}
We now use boost arguments first used in \cite{Wall} and later in \cite{SPANMS}, \cite{BBDK}.
The metric near any null hypersurface and therefore near the event horizon can be given in Gaussian null coordinates as
\begin{equation}\label{I8}
ds^2= 2dv du -u^2 X(u,v,x^{k}) dv^2 + 2 u \omega_{i}(u,v,x^{k}) dv dx^{i} + h_{ij}(u,v,x^{k})dx^{i} dx^{j}
\end{equation}
where $v$ is an affine parameter along the null generator of the horizon, $x^{i}$ corresponds to coordinates on the $D-2$ transverse surface (cut) and $u$ is chosen in a way that $\partial_v. \partial_u=1$ and $\partial_u. \partial_i=0$. In this coordinate system, $u=0$ is the future horizon and $u=0, v=0$ corresponds to the bifurcation surface $\mathcal{B}$. These coordinates may not cover the entire spacetime, but the near-horizon region of any dynamical black hole spacetime can always be written in this form. Now, the black hole spacetime we consider in this paper is a static (therefore stationary) black hole spacetime which is a solution in a diffeomorphism invariant theory of gravity. Then, the black hole event horizon is a Killing horizon \cite{Wald}. First consider the case where this horizon is a Killing horizon with respect to the boost field $\xi= v\partial_v - u\partial_u $ (see \cite{SPANMS}). This is true for any stationary black hole spacetime which, near the horizon, looks like a Rindler spacetime, hence the terminology `boost field'. The near-horizon metric of this stationary black hole will then be boost invariant, i.e the Lie derivative of the metric $\mathcal{L}_{\xi} g_{\mu \nu}=0$.
Then, the near-horizon metric (\ref{I8}) is of the form
\begin{equation}\label{I9}
ds^2_{st}= 2dv du -u^2 X(uv,x^{k}) dv^2 + h_{ij}(uv,x^{k})dx^{i} dx^{j}
\end{equation}
Here, $\omega_{i} = 0$ since the spacetime is static.
This is the most general form of a static spacetime with a Killing horizon near the horizon.
It can easily be seen from (\ref{I8}) that along the horizon, any non-zero tensor $A$, which is constructed out of metric components can always be written as $A=\partial^n_v \partial^m_u B$, where $m,n$ are integers and $B$ is constructed out of metric components $X,\omega_i,h_{ij} $ and their derivatives with respect to $\nabla_i$ . Then, we can associate a boost weight with these tensors as boost weight $=$ $\#v$ index $-$ $\#u$ index. Furthermore, we can write the schematic form  for the $vv$ component of any 2-tensor $A_{vv}$ constructed from metric components as
\begin{equation}\label{I10}
A_{vv}= \tilde X\partial_{v}^2 Y + C\partial_v A \partial_v B .
\end{equation}
Here,  $\tilde X,Y,C,A,B$ have boost weight zero and are constructed out of metric components. Now for the stationary black hole spacetime, the above equation reduces down to
\begin{equation}\label{I11}
A_{vv}|_{st}= u^2 \tilde X\partial_{uv}^2 Y + u^2 C\partial_{uv} A \partial_{uv} B .
\end{equation}
This is because the stationary black hole has a Killing symmetry which reduces on the horizon to a scaling symmetry under $u \to p u$ and $v \to v/p$. Thus, the metric components in the stationary case only depend on $uv$ at the horizon. This implies that the $vv$ component of any 2-tensor $A_{vv}$ constructed from metric components in a stationary black hole spacetime vanishes at the future horizon i.e at $u=0$.
\begin{equation}\label{I12}
A_{vv}|_{st}^{u=0}=0.
\end{equation}
Now, the $vv$ component of the equation of motion for any higher curvature theory takes the following form,
\begin{equation}\label{I13}
R_{vv} + H_{vv}= 8\pi T_{vv}
\end{equation}
where $H_{vv}$ corresponds to a higher curvature contribution to the equation of motion. Using (\ref{I11}) for the stationary black hole $R_{vv}=0$ and $H_{vv}=0$, this implies that
\begin{equation}\label{I14}
T_{vv}|^{u=0}=0
\end{equation}
for any \emph{classical} matter stress-energy tensor. Furthermore, whenever a $v$ derivative acts on the stationary background metric component, it gives a factor of $u$ as well, since the metric component depends on $v$ only through $uv$. Hence such a term will vanish at the future horizon $u=0$. Therefore, from (\ref{I10}), the $vv$ component of any 2- tensor $A_{vv}$ \emph{linear} in the metric perturbation at the future horizon $u=0$ can always be written in the following form,
\begin{equation}\label{I15}
A_{vv}|^{u=0}= \partial_{v}^2 \zeta ,
\end{equation}
where $\zeta$ has boost weight zero and is constructed from the background metric and the linear perturbation over stationarity.
\\
\subsection{Semi-classical gravity equations }
Following Chandrasekaran, Penington and Witten \cite{VGE} and Wall \cite{AW}, we will work in semi-classical gravity where the expectation value of the matter stress energy tensor is a source term in the gravity equations.
Now let us look at the semiclassical equations of motion. The $vv$ component is
\begin{equation}\label{I16}
R_{vv}+ H_{vv}= 8\pi <T_{vv}>
\end{equation}
Writing this order by order in $\epsilon$, we get:
\begin{eqnarray}
\epsilon^0: \hspace{25mm} R_{vv}+H_{vv}=0   \label{I17} \\
\epsilon^{\frac{1}{2}}: \hspace{22mm} R^{(\frac{1}{2})}_{vv}+H^{(\frac{1}{2})}_{vv}=0  \label{I18}\\
\epsilon^{1} :\hspace{5mm} R^{(1)}_{vv}+ H^{(1)}_{vv}= 8\pi <T^Q_{vv}>.  \label{I19}
\end{eqnarray}
where $T^Q_{vv}=T^M_{vv}+t_{vv}$, $<T^M_{vv}>\hspace{2mm} \sim O(\epsilon)$ is a matter stress energy tensor and $<t_{vv}> \hspace{2mm} \sim O(\epsilon)$, the pseudo-stress-energy tensor of the graviton.
Further, $R^{(\frac{1}{2})}_{vv}$, $H^{(\frac{1}{2})}_{vv}$ are linear in $g_{\mu \nu}^{(\frac{1}{2})}$ perturbation, and $R^{(1)}_{vv}$, $H^{(1)}_{vv}$ are linear in $g_{\mu \nu}^{(1)}$ perturbation.
\subsection{Raychaudhuri equation order by order}
As is well-known, the Raychaudhuri equation (\ref{I18}) plays a key role in the proof of the second law and we will use it later in our computation. The Raychaudhuri equation is given by,
\begin{equation}\label{I20}
\frac{d\theta}{dv}=-\Big\{\frac{\theta^2}{D-2} + \sigma^{\alpha \beta} \sigma_{\alpha \beta} + R_{vv} \Big\}
\end{equation}
Now, if we write it order by order in $\epsilon$, we get,
\begin{eqnarray}
\epsilon^{0}: \hspace{71mm}   \frac{d\theta^{(0)}}{dv}=0 \label{I21}\\
\epsilon^{\frac{1}{2}}:\hspace{61mm} \frac{d\theta^{(\frac{1}{2})}}{dv}= -R^{(\frac{1}{2})}_{vv} \label{I22}\\
\epsilon^1 :  \hspace{2mm}  \frac{d\theta^{(1)}}{dv}=-\Big\{\frac{\theta^{(\frac{1}{2})} \theta^{(\frac{1}{2})}}{D-2} + \sigma^{\alpha \beta}_{(\frac{1}{2})} \sigma_{\alpha \beta}^{(\frac{1}{2})} +R^{(1)}_{vv}  + R^{(\frac{1}{2},\frac{1}{2})}_{vv}\Big\}\label{I23}
\end{eqnarray}
Further using (\ref{I19}), (\ref{I23}) can be written as
\begin{equation}\label{I24a}
\frac{d\theta^{(1)}}{dv}=-\Big\{\frac{\theta^{(\frac{1}{2})} \theta^{(\frac{1}{2})}}{D-2} + \sigma^{\alpha \beta}_{(\frac{1}{2})} \sigma_{\alpha \beta}^{(\frac{1}{2})} +8\pi <T^Q_{vv}> - H^{(1)}_{vv} + R^{(\frac{1}{2},\frac{1}{2})}_{vv}\Big\}.
\end{equation}
Furthermore, if we compute $R_{vv}^{(\frac{1}{2},\frac{1}{2})}$  in TT (transverse traceless) gauge at the horizon \footnote{The perturbative expansion of the Ricci tensor to quadratic order can be found in \cite{MTW}.}, we will get,
\begin{equation}\label{I24a1}
R^{(\frac{1}{2},\frac{1}{2})}_{vv} \underset{TT}{=}-\frac{1}{4}\frac{d g^{ij}_{(\frac{1}{2})}}{dv}\frac{d g_{ij}^{(\frac{1}{2})}}{dv}+ \frac{1}{2} \frac{d}{dv}\Big( g^{ij}_{(\frac{1}{2})}\frac{d g_{ij}^{(\frac{1}{2})}}{dv}\Big).
\end{equation}
Now using the fact that $\frac{1}{2}\frac{d g_{ij}^{(\frac{1}{2})}}{dv}= \sigma^{(\frac{1}{2})}_{ij}+\frac{1}{D-2}g^{(0)}_{ij}\theta^{(\frac{1}{2})}$ \cite{AW} ($i,j$ are transverse coordinate indices), we can write equation (\ref{I24a1}) as
\begin{equation}\label{I24b}
R^{(\frac{1}{2},\frac{1}{2})}_{vv} \underset{TT}{=} - \Big(\frac{\theta^{(\frac{1}{2})} \theta^{(\frac{1}{2})}}{D-2} + \sigma^{\alpha \beta}_{(\frac{1}{2})} \sigma_{\alpha \beta}^{(\frac{1}{2})}\Big) +\frac{1}{4}\frac{d^2}{dv^2}\Big(g^{ij}_{(\frac{1}{2})}g_{ij}^{(\frac{1}{2})}\Big)
\end{equation}
which yields
\begin{equation}\label{I25}
\frac{d\theta^{(1)}}{dv}=-\Big\{ 8\pi <T^Q_{vv}> - H^{(1)}_{vv} +\frac{1}{4}\frac{d^2}{dv^2}\Big(g^{ij}_{(\frac{1}{2})}g_{ij}^{(\frac{1}{2})}\Big) \Big\}.
\end{equation}
(\ref{I21}) follows from the fact that the background solution is stationary. The other equations are obtained by expanding the Raychaudhuri equation order by order.
\subsection{Entropy change due to accretion of quantum matter across the horizon}
In this subsection, we will compute the order-by-order change in the entropy due to the accretion of quantum matter and gravitons across the horizon. We will assume that the background black hole solution is stationary, as well as the final state of the black hole at late times. Also, we assume the perturbation will fall off fast enough, so that all boundary terms at late times vanish \footnote{This implies that at late times all the perturbations would have either crossed the horizon or gone to asymptotic infinity. For AdS black hole spacetime with reflecting boundary conditions, all perturbations will cross the horizon. Also, we dynamically impose the gauge (\ref{I8}) at all times. }. To compute the entropy change order by order, first we will write the perturbative expansion of entropy density as
\begin{equation}\label{I26}
\rho= \rho^{(0)}+ \epsilon^{\frac{1}{2}} \rho^{(\frac{1}{2})}+ \epsilon \rho^{(1)}+ O(\epsilon^{\frac{3}{2}} )
\end{equation}
Now, we have all the tools to compute the change in entropy. First we will do the change in entropy computation in the absence of the graviton. Then we will do the computation in which we will include gravitons.
\subsubsection {Entropy change without graviton contribution}\label{EAG}
When there is no graviton, all the terms with $g_{\mu \nu}^{(\frac{1}{2})}$ perturbation will go away in all of the above equations. Also, the stress-energy tensor will have only the matter contribution i.e, $<T^{Q}_{vv}>= <T^{M}_{vv} > $, which we will take to be $O(\epsilon)$. The background solution is a stationary black hole, with a Killing horizon and regular bifurcation sphere. For the stationary black hole, the expansion coefficient of the horizon is zero and the entropy density will be independent of the chosen horizon cut.  Therefore
\begin{equation}\label{EAG1}
\epsilon^0: \hspace{5mm} \Delta S^{(0)}=0
\end{equation}
Now, we will compute the change in the entropy due to accretion of matter by the stationary black hole, which takes it away from stationarity. As we have already mentioned, the black hole will settle down into a new stationary state at late times.
 Now using (\ref{I6}) and (\ref{I25}) with the fact that there is no $\rho^{(\frac{1}{2})}$  and $\Theta^{\frac{1}{2}}$ we will get
\begin{equation}\label{EAG2}
\epsilon: \hspace{5mm} \Delta \delta S^{(1)}= -\frac{1}{4}\int_{0}^{\infty} dv \int  d^{D-2}x \sqrt{h} v \Big\{ \frac{d^2 \rho^{(1)} }{dv^2} + \frac{d\theta^{(1)}}{dv} \rho^{(0)}  \Big\}.
\end{equation}
Here, $\delta$ is the perturbation away from stationarity.
Since $\rho^{(0)} = 1 +\rho^{(0)}_w $,
\begin{equation}\label{EAG2a}
\frac{d\theta^{(1)}}{dv}\rho^{(0)}= \frac{d\theta^{(1)}}{dv}+\frac{d\theta^{(1)}}{dv}\rho^{(0)}_w =-8\pi<T_{vv}^{Q}>+H_{vv}^{(1)} -R_{vv}^{(1)}\rho^{(0)}_w .
\end{equation}
The equation (\ref{EAG2a}) is obtained using (\ref{I23}),(\ref{I25}) and the fact that there is no $g^{(\frac{1}{2})}$ perturbation, which will make $\frac{d\theta^{(1)}}{dv}=-R_{vv}^{(1)}$. Further, we used the equation of motion to rewrite $R_{vv}^{(1)}$ in terms of the stress energy tensor. Putting the equation (\ref{EAG2a}) in the equation (\ref{EAG2}), we get
\begin{equation}\label{EAG3}
\epsilon: \hspace{5mm} \Delta \delta S^{(1)}= -\frac{1}{4}\int_{0}^{\infty} dv \int  d^{D-2}x \sqrt{h} v \Big\{ \Big(\frac{d^2 \rho^{(1)} }{dv^2} - R^{(1)}_{vv} \rho^{(0)}_{w} + H^{(1)}_{vv}\Big) -8\pi<T^{Q}_{vv}>  \Big\}.
\end{equation}
We note that $\Big(\frac{d^2 \rho^{(1)} }{dv^2} - R^{(1)}_{vv} \rho_{w}^{(0)} + H^{(1)}_{vv}\Big) $ is constructed out of background metric components and the perturbation and is linear in the perturbation \footnote{We can replace the ordinary derivatives with respect to $v$ in the first term with covariant derivatives in the gauge we are in.}. Therefore, using boost arguments we can write $\Big(\frac{d^2 \rho^{(1)} }{dv^2} - R^{(1)}_{vv} \rho_{w}^{(0)} + H^{(1)}_{vv}\Big)= \partial^{2}_{v} \zeta_{(1)} $, which will yield
\begin{equation}\label{EAG4}
 \Delta \delta S^{(1)}= -\frac{1}{4}\int_{0}^{\infty} dv \int  d^{D-2}x \sqrt{h} v \Big\{ \partial^{2}_{v} \zeta_{(1)} -8\pi<T^{Q}_{vv}>  \Big\}.
\end{equation}
We can simplify the above term using integration by parts,
\begin{multline}\label{EAG5}
 \Delta \delta S^{(1)}= 2\pi \int_{0}^{\infty} dv \int  d^{D-2}x \sqrt{h} v <T^{Q}_{vv}> +\frac{1}{4}\int   d^{D-2}x \sqrt{h} \int_{0}^{\infty} dv \partial_{v} \zeta_{(1)}\\ - \frac{1}{4}\int  d^{D-2}x \sqrt{h} \Big(v \partial_{v} \zeta_{(1)}\Big)\Big |_{v=0} ^{v\rightarrow \infty}.
\end{multline}
Now we assume fall-off conditions at late times i.e, all perturbations and their derivatives should fall fast enough such that this boundary term goes to zero at late times. The contribution from the last term in (\ref{EAG5}) also vanishes at $v=0$. Let us recall that we are in the gauge (\ref{I8}) in which the horizon is always at $u=0$. Then we will get
\begin{equation}\label{EAG6}
\Delta \delta S^{(1)}= 2\pi \int_{0}^{\infty} dv \int  d^{D-2}x \sqrt{h} v <T^{Q}_{vv}>  - \frac{1}{4}\int   d^{D-2}x \sqrt{h}  \zeta_{(1)}\Big |_{v=0}.
\end{equation}
We can get rid of the second term by assuming that we can fix the JKM ambiguity in such a way that
\begin{equation}\label{EAG7}
\frac{d^2 \rho^{(1)} }{dv^2} - R^{(1)}_{vv} \rho_{w}^{(0)} + H^{(1)}_{vv}= \partial^2 _{v} \zeta_{(1)} =0
\end{equation}
  everywhere on the horizon. This will get rid of the term which is giving rise to the second term in (\ref{EAG6}). This is because then $\zeta_{(1)}(v)= a v + b$, where $a$ and $b$ are only functions of transverse coordinate. $\zeta_{(1)}$ is constructed out of the background metric and the linear perturbation in the gauge (\ref{I8}). We have to further impose the fall-off conditions on the perturbation i.e., the perturbation and its derivatives with respect to $v$ must go to zero at late times. Thus $\zeta_{(1)}(v)=0$. There is no contradiction with the fact that $\Omega$ vanishes at the bifurcation surface in linear order. It is shown in \cite{ASAS} that the second term in (\ref{EAG6}) is zero for $F(R)$ gravity and arbitrary order Lovelock gravity. It is also argued there that this will be true for an arbitrary diffeomorphic theory at linear order. Therefore, assuming this,
\begin{equation}\label{EAG8}
\Delta \delta S^{(1)}= 2\pi \int_{0}^{\infty} dv \int  d^{D-2}x \sqrt{h} v <T^{Q}_{vv}>
\end{equation}
The above derivation is of course true even when the accreting matter is classical. For classical matter, imposing the null energy condition i.e $T^{Q}_{vv} \geq 0$ will give the second law.

\subsubsection {Entropy change with the graviton contribution included}
In this section, we include the graviton contribution, and therefore we will work with the full perturbation expansion. We will again do an order by order expansion. Using (\ref{I6}) and the fact that background solution is stationary,
\begin{equation}\label{I26a}
\epsilon^0: \hspace{5mm} \Delta S^{(0)}=0
\end{equation}
 Now, let us compute change in entropy at $\epsilon^{\frac{1}{2}}$ order. Writing (\ref{I6}) at $\epsilon^{\frac{1}{2}}$ will give
\begin{equation}\label{I27}
\epsilon^{\frac{1}{2}}: \hspace{5mm} \Delta \delta S^{(\frac{1}{2})}= -\frac{1}{4}\int_{0}^{\infty} dv \int  d^{D-2}x \sqrt{h} v \Big\{ \frac{d^2 \rho^{(\frac{1}{2})} }{dv^2} + \frac{d\theta^{(\frac{1}{2})}}{dv} \rho^{(0)}  \Big\}.
\end{equation}
Here, $\delta$ corresponds to entropy change due to a perturbation that takes the solution away from stationarity. Using (\ref{I22}), we can write (\ref{I27}) as
\begin{equation}\label{I27a}
\Delta \delta S^{(\frac{1}{2})}= -\frac{1}{4}\int_{0}^{\infty} dv \int  d^{D-2}x \sqrt{h} v \Big\{ \frac{d^2 \rho^{(\frac{1}{2})} }{dv^2} - R_{vv}^{(\frac{1}{2})} \rho^{(0)}  \Big\}
\end{equation}

Furthermore using the boost argument in (\ref{I15}), we can write $R_{vv}^{(\frac{1}{2})}= \partial_v^2 \zeta_{(\frac{1}{2})}$, where $\zeta_{(\frac{1}{2})}$ is constructed out of the background metric and linear perturbation in $g_{\mu \nu}^{(\frac{1}{2})}$ where we work in the gauge (\ref{I8}). Further, $\rho^{(0)}$ is independent of $v$. Therefore, the equation (\ref{I27a}) becomes
\begin{equation}\label{I28}
\Delta \delta S^{(\frac{1}{2})}= -\frac{1}{4}\int_{0}^{\infty} dv \int  d^{D-2}x \sqrt{h} v \frac{d^2}{dv^2}\Big\{ \rho^{(\frac{1}{2})}- \rho^{(0)} \zeta_{(\frac{1}{2})}\Big\}.
\end{equation}
Using integration by parts,
\begin{equation}\label{I29}
\Delta \delta S^{(\frac{1}{2})}= -\frac{1}{4}\int d^{D-2}x \sqrt{h} v \frac{d}{dv}\Big\{ \rho^{(\frac{1}{2})}- \rho^{(0)} \zeta_{(\frac{1}{2})}\Big\} \Big|_{v=0}^{v \rightarrow \infty}
+\frac{1}{4}\int_{0}^{\infty} dv \int  d^{D-2}x \sqrt{h}  \frac{d}{dv}\Big\{ \rho^{(\frac{1}{2})}- \rho^{(0)} \zeta_{(\frac{1}{2})}\Big\}.
\end{equation}
Using the fall-off condition on the perturbation at late times, the first term vanishes. Therefore, we get
\begin{equation}\label{I30}
\Delta \delta S^{(\frac{1}{2})}=\frac{1}{4}\int_{0}^{\infty} dv \int  d^{D-2}x \sqrt{h}  \frac{d}{dv}\Big\{ \rho^{(\frac{1}{2})}- \rho^{(0)} \zeta_{(\frac{1}{2})}\Big\}.
\end{equation}
Integrating and using fall-off condition gives
\begin{equation}\label{I31}
\Delta \delta S^{(\frac{1}{2})}=-\frac{1}{4} \int  d^{D-2}x \sqrt{h} \Big\{ \rho^{(\frac{1}{2})}- \rho^{(0)} \zeta_{(\frac{1}{2})}\Big\}\Big|_{v=0}.
\end{equation}
Using the fact that $\theta=\frac{d }{dv} \log{(\sqrt{h})}$ and (\ref{I22}), we can write $\delta \log {(\sqrt{h})}=-\zeta_{(\frac{1}{2})}$. Therefore, we can write (\ref{I31}) as
\begin{equation}\label{I32}
\Delta \delta S^{(\frac{1}{2})}=-\frac{1}{4} \int_{\mathcal{B}}  d^{D-2}x \sqrt{h} \Big\{ \rho^{(\frac{1}{2})}+ \delta \log {(\sqrt{h})} \rho^{(0)}\Big\}.
\end{equation}
Using (\ref{I1}), it is straightforward to verify that
\begin{equation}\label{I33}
\Delta \delta S^{(\frac{1}{2})}=-\delta S^{(\frac{1}{2})}\Big|_{\mathcal{B}}
\end{equation}
$\mathcal{B}$ is the bifurcation surface.
Now, we use Theorem $6.1$ in Iyer and Wald (IW) \cite{VR}, i.e  $(\delta S= \delta \mathcal{E}- \Omega_{H} \delta \mathcal{J})\Big|_{\mathcal{B}}$, where $\mathcal{E}$ is the canonical energy and $\mathcal{J}$ is the canonical angular momentum of the black hole in the covariant phase space formalism \footnote{Since we have a static black hole, angular velocity at the horizon is zero.}. This was proved by IW at the bifurcation surface for any non-stationary perturbation satisfying the linearized equation of motion\footnote{IW's first law at the bifurcation surface $\mathcal{B}$ is unaffected by the JKM ambiguity.}. Now in our case, there is no stress-energy tensor at $\epsilon^{(\frac{1}{2})}$ order. This implies
\begin{equation}\label{I34}
\Delta \delta S^{(\frac{1}{2})}=-\delta S^{(\frac{1}{2})}\Big|_{\mathcal{B}}=0
\end{equation}
Hence, $\delta S^{(\frac{1}{2})}(0)= \delta S^{(\frac{1}{2})}(\infty)=0$. Therefore, from (\ref{I34}), if $\delta S^{(\frac{1}{2})}$ is non-zero at any cut, then for some range of $v$ entropy will definitely decrease. This violates the second law. The only way for the second law to be true is to assume that $\delta S^{(\frac{1}{2})}$ will vanish at arbitrary cut. It was shown explicitly by the authors in \cite{AS,ASAS} that
\begin{equation}\label{I35}
\mathcal{R}_{vv}\equiv \frac{d^2\rho}{dv^2}- R_{vv} \rho_{w}^{(0)} + H_{vv}
\end{equation}
 vanishes for $F(R)$ theory and Lovelock theory of arbitrary order, at the linear order in perturbation theory about the stationary black hole (perturbation can be non-stationary). Using (\ref{I18}), it can  can be checked that the term in curly brackets in (\ref{I27}) is the same as $\mathcal{R}_{vv}^{(\frac{1}{2})}$. The authors in \cite{AS, ASAS} also argued that this relation may be true for an arbitrary theory of gravity with an appropriate definition of local entropy density. Vanishing of $\delta S^{(\frac{1}{2})}$ in general will yield,
 \begin{equation}\label{I36}
 \frac{d^2\rho^{(\frac{1}{2})}}{dv^2}= - \rho^{(0)} \partial_v \theta_{(\frac{1}{2})}
\end{equation}
which after integration and using the boundary condition that the perturbation vanishes at late times will give  $\rho^{(\frac{1}{2})}=\rho^{(0)} \zeta_{(\frac{1}{2})} $.
\\
 Now, let us compute the $\epsilon$ order change in entropy, writing (\ref{I6}) to the $\epsilon$ order. We get
 \begin{equation}\label{I37}
 \Delta \delta S^{(1)}=-\frac{1}{4}\int_{0}^{\infty}dv \int v \Big\{\frac{d^2\rho^{(1)} }{dv^2}+ \rho^{(0)}\frac{d\theta^{(1)}}{dv} +\frac{d\theta^{(\frac{1}{2})}}{dv} \rho^{(\frac{1}{2})} + 2 \theta^{(\frac{1}{2})} \frac{d\rho^{(\frac{1}{2})}}{dv} + \rho^{(0)} \theta^{(\frac{1}{2})} \theta^{(\frac{1}{2})} \Big\} \sqrt{h} d^{D-2}x.
\end{equation}
Using (\ref{I19}), (\ref{I23}) and (\ref{I35}), the first two terms in the above expression can be written as
\begin{equation}\label{I38}
\frac{d^2\rho^{(1)} }{dv^2}+ \rho^{(0)} \frac{d\theta^{(1)}}{dv} =-8\pi<T^{Q}_{vv}> + \mathcal{R}^{(1)}_{vv}-\frac{1}{4}\frac{d^2}{dv^2}\Big(g^{ij}_{(\frac{1}{2})}g_{ij}^{(\frac{1}{2})}\rho^{(0)}\Big).
\end{equation}
Using (\ref{I35}), (\ref{I15}) and the fact that $H^{(1)}_{vv}$ and $R^{(1)}_{vv}$ are constructed out of background metric components and the perturbation and are linear in $g^{(1)}_{\mu \nu}$ perturbation in the gauge (\ref{I8}), $\mathcal{R}^{(1)}_{vv}$ can be written as $\mathcal{R}^{(1)}_{vv}= \partial_{v}^2 \zeta_{(1)}$.  This yields
\begin{equation}\label{I38a}
\frac{d^2\rho^{(1)} }{dv^2}+ \rho^{(0)} \frac{d\theta^{(1)}}{dv} =-8\pi<T^{Q}_{vv}> + \partial_v^2 \zeta'_{(1)}.
\end{equation}
where, $\zeta'_{(1)}=\zeta_{(1)}-\frac{1}{4}\Big(g^{ij}_{(\frac{1}{2})}g_{ij}^{(\frac{1}{2})}\rho^{(0)}\Big)$.Putting the above equation in (\ref{I37}) we get,
\begin{multline}\label{I39}
-\frac{1}{4}\int_{0}^{\infty}dv \int v \sqrt{h} d^{D-2}x \Big(\frac{d^2\rho^{(1)} }{dv^2}+ \rho^{(0)} \frac{d\theta^{(1)}}{dv}\Big) =2\pi\int_{0}^{\infty}dv \int v \sqrt{h} d^{D-2}x   <T^{Q}_{vv}>  \\  - \frac{1}{4} \int \sqrt{h} \zeta'_{(1)} d^{D-2}x \Big|_{v=0}.
\end{multline}
We get the above equation using integration by parts in the $\partial_v^2 \zeta'_{(1)}$ integral and the fact that the term at $v\rightarrow \infty$ will vanish due to the fall-off condition. Let us consider the rest of the terms in (\ref{I37}), we will call it
$A(\frac{1}{2},\frac{1}{2})$,
\begin{equation}\label{I40}
A(\frac{1}{2},\frac{1}{2}) = -\frac{1}{4}\int_{0}^{\infty}dv \int v \Big\{\frac{d\theta^{(\frac{1}{2})}}{dv} \rho^{(\frac{1}{2})} + 2 \theta^{(\frac{1}{2})} \frac{d\rho^{(\frac{1}{2})}}{dv} + \rho^{(0)} \theta^{(\frac{1}{2})} \theta^{(\frac{1}{2})} \Big\} \sqrt{h} d^{D-2}x.
\end{equation}
After integrating (\ref{I36}) once, we get $\frac{d\rho^{(\frac{1}{2})}}{dv}=-\rho^{(0)} \theta_{(\frac{1}{2})}$.  Putting this in (\ref{I40}) will yield
\begin{equation}\label{I41}
A(\frac{1}{2},\frac{1}{2}) = -\frac{1}{4}\int_{0}^{\infty}dv \int v \Big\{\frac{d\theta^{(\frac{1}{2})}}{dv} \rho^{(\frac{1}{2})} +  \theta^{(\frac{1}{2})} \frac{d\rho^{(\frac{1}{2})}}{dv}  \Big\} \sqrt{h} d^{D-2}x
\end{equation}
which can be further simplified using integration by parts and using fall-off conditions as $v \to \infty$,
\begin{equation}\label{I42}
A(\frac{1}{2},\frac{1}{2}) = \frac{1}{4}\int_{0}^{\infty}dv \int \theta^{(\frac{1}{2})} \rho^{(\frac{1}{2})} \sqrt{h} d^{D-2}x .
\end{equation}
Using $\frac{d\rho^{(\frac{1}{2})}}{dv}=-\rho^{(0)} \theta_{(\frac{1}{2})}$ in (\ref{I42}), we get
\begin{equation}\label{I43}
A(\frac{1}{2},\frac{1}{2}) = -\frac{1}{8}\int_{0}^{\infty}dv \int  \frac{d}{dv} \Big(\frac{(\rho^{(\frac{1}{2})})^2}{\rho^{(0)}}\Big) \sqrt{h} d^{D-2}x.
\end{equation}
After integrating and using fall-off at late times, we will get
\begin{equation}\label{I44}
A(\frac{1}{2},\frac{1}{2}) = \frac{1}{8} \int  \Big(\frac{(\rho^{(\frac{1}{2})})^2}{\rho^{(0)}}\Big) \sqrt{h} \Big|_{v=0} d^{D-2}x
\end{equation}
This quantity is thus manifestly positive. From (\ref{I39}) and (\ref{I44}),  we get
\begin{multline}\label{I45}
\Delta \delta S^{(1)}= 2\pi\int_{0}^{\infty}dv \int v \sqrt{h} d^{D-2}x   <T^{Q}_{vv}>
- \frac{1}{4} \int \sqrt{h} \zeta'_{(1)} d^{D-2}x \Big|_{v=0}  + A(\frac{1}{2},\frac{1}{2}).
\end{multline}
We know that the Wald entropy has JKM ambiguities when the metric is not stationary. That was a motivation for putting $\Omega$  as the correction to the Wald entropy in the definition of local entropy density. We now fix $\Omega$ such that the last two terms in (\ref{I45}) vanish. These terms are anyway zero for a stationary black hole, as can be seen from the expression for $A(\frac{1}{2},\frac{1}{2})$ in (\ref{I44}) and $\mathcal{R}^{(1)}_{vv}$ is zero for a stationary black hole from results in the subsection on the boost argument \footnote{For a stationary black hole, $v$ derivatives of $\rho$ are zero and $R_{vv}$ and $H_{vv}$ are zero for a stationary metric as discussed using boost arguments.}. So $\Omega$ will be non-zero only when the metric is not stationary. For the cases when it is possible to set these two terms to zero by a choice of $\Omega$, we will get
\begin{equation}\label{I46}
\Delta \delta S^{(1)}= 2\pi\int_{0}^{\infty}dv \int v \sqrt{h} d^{D-2}x   <T^{Q}_{vv}>
\end{equation}
Some of the ambiguities in Wald's entropy were fixed for some class of theories at linear order in perturbation theory\cite{Wall}. Moreover, this entropy was shown to be equal to holographic entanglement entropy computed by Dong \cite{Dong}.It is also pointed out in \cite{Wall} that considering the second law at linear order does not fix the ambiguities at higher order. Therefore these results are not in contradiction with our computation.  One can also get rid of the second and third term in (\ref{I45}), by restricting the perturbation to a special class of perturbations which vanishes at $v=0$. One physical case in which such perturbation can be realised is when matter falls after $v=0$.
\section {The Entropy of Algebra And Generalized Entropy in Higher Curvature Theory}
In this section, we utilize the algebraic approach to quantum field theory, specifically the constructions of Chandrasekaran, Longo,  Penington, and Witten (CLPW) in \cite{VRGE}, and Chandrasekaran, Penington, and Witten (CPW) in \cite{VGE}, to study black holes in higher curvature theories. We have employed their construction to prove a version of the GSL (Generalized Second Law) for an arbitrary diffeomorphism invariant theory of gravity, with certain appropriate assumptions.
The setup that we are interested in involves both the asymptotically flat and asymptotically AdS stationary black hole solutions in an arbitrary diffeomorphism invariant theory. Throughout this section we follow the notation of CPW. In this section, we will first review the work of CPW and then generalize it to our case.
\subsection{Brief review of the recent papers }
  The series of recent papers by CLPW, Witten and CPW \cite{VRGE, EW3, VGE} have helped to understand the generalized entropy introduced by Bekenstein \cite{JB} better. They have addressed the question of why the generalized entropy is well-defined, whereas the gravity contribution and the quantum field contribution in the generalized entropy are not well-defined separately. CPW showed for an eternal black hole that is either asymptotically flat and asymptotically AdS, that there is an entropy associated with the von Neumann algebra of quantum fields (including the graviton) in the black hole left or right exterior, together with either the left or right ADM/boundary Hamiltonian. Further, they showed that this algebra entropy is the generalized entropy at the bifurcation surface of the black hole. They have also discussed the monotonicity of the generalized entropy of asymptotically AdS black holes by using techniques from von Neumann algebras.

  We first provide a brief review of the salient features of these constructions and then generalize them to our case of black holes in arbitrary diffeomorphism invariant theories of gravity.

\begin{figure}[h]
  \centering
  \includegraphics[width=0.80\textwidth]{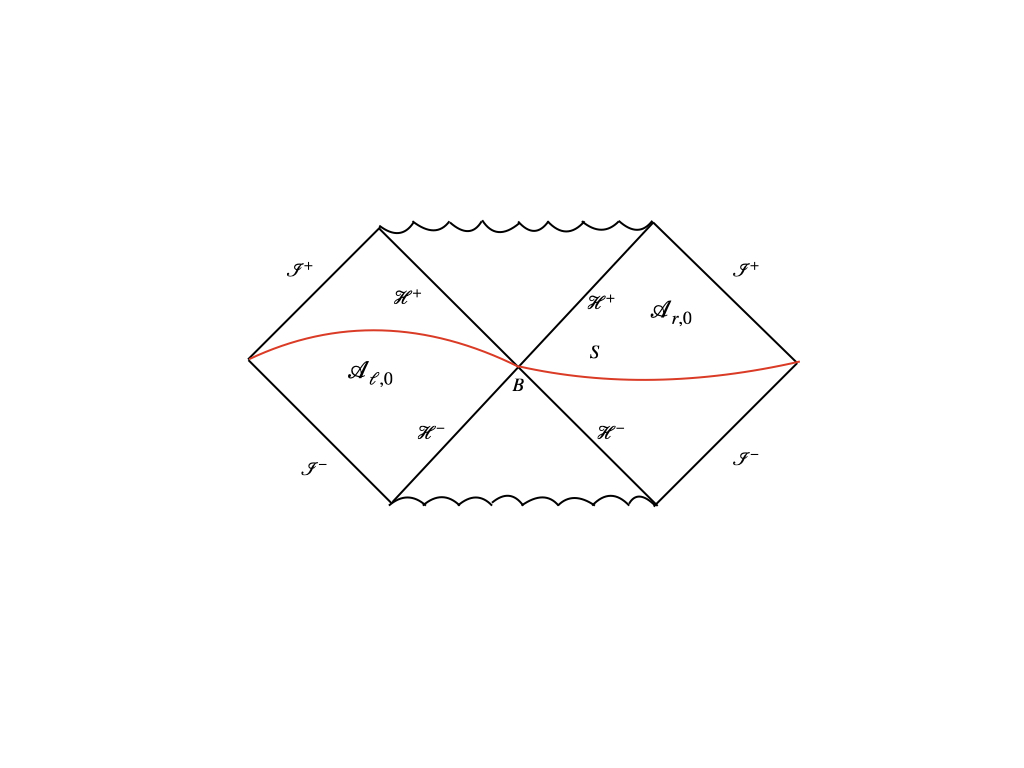}
  \caption{Maximally extended stationary black hole with Cauchy surface $S$. The bulk algebra of the left and the right exterior region is $\mathcal{A}_{\ell,0}$ and $\mathcal{A}_{r,0}$ respectively.}
  \label{fig:BH2}
\end{figure}
Let $M$ be the asymptotically flat, maximally extended Schwarzschild black hole in Einstein's theory or the maximally extended AdS-Schwarzschild black hole. We consider quantum fields in this spacetime, including gravitons. The left and right exterior regions of $M$  will be denoted by $\ell$ and $r$ respectively, while $L$ and $R$ will be used to denote left and right spatial infinity. Let $\mathcal{H}_{0}$ be the Hilbert space of this theory that we get by quantizing the fields and the local algebra of observables of the left and right exterior region be $\mathcal{A}_{\ell,0}$ and $\mathcal{A}_{r,0}$ respectively as shown in Figure \ref{fig:BH2}. It is well known that algebras $\mathcal{A}_{\ell,0}$ and $\mathcal{A}_{r,0}$   are Type  \RomanNumeralCaps {3}$_1$ factors (their centers are trivial) \cite{HA, EW1,EW2}\footnote{The algebra of operators in quantum field theory in a causal wedge is always a von Neumann algebra of Type \RomanNumeralCaps {3} \cite{LL}. When the center is trivial it is a Type \RomanNumeralCaps {3}$_1$ algebra. }. Moreover, $\mathcal{A}_{\ell,0}$ and $\mathcal{A}_{r,0}$ are each other's commutants (i.e all the operators of $\mathcal{A}_{\ell,0}$ commute with all the operators in $\mathcal{A}_{r,0}$).
\\
 The spacetime is stationary and equipped with a time translation Killing field $V$. $V$ is future directed in the right exterior region and past-directed in the left exterior region. Due to background diffeomorphism invariance, one can define a conserved quantity $\hat{h}$ associated with the time translation vector field $V$. Let $S$ be the bulk Cauchy surface going from the spatial infinity of the right exterior region to the spatial infinity of the left exterior region, through the bifurcation surface as shown in Figure \ref{fig:BH2}. Then $\hat{h}$ can be defined as
 \begin{equation}\label{II1}
 \hat{h}=\int_{S}d\Sigma^{\mu}V^{\nu} T_{\mu \nu} .
 \end{equation}
  Here, $T_{\mu \nu}$ is the stress-energy tensor of the bulk fields\footnote{$T_{\mu \nu }$ includes the contribution from the pseudo-stress tensor of gravitons.}. In Tomita-Takesaki theory of the quantum fields in the black hole exterior, $\beta \hat{h}$ is the modular Hamiltonian associated with the Hartle-Hawking state $|\Psi_{HH}>$ of the black hole and $\beta$ is the Hawking temperature\footnote{The modular operator is defined as $\Delta=\exp{(-H_{mod})}$ and for $\ket{\Psi_{HH}}$, $H_{mod}=\beta \hat{h}$.}\cite{JE, GS}. It is well known that $\hat{h}$ in Einstein's gravity is the difference between the right ADM Hamiltonian $H_{R}$ and the left ADM Hamiltonian $H_{L}$ i.e. $\hat{h}= H_{R}- H_{L}$.

  CLPW and CPW now extend the Type \RomanNumeralCaps {3}$_1$ algebra $\mathcal{A}_{\ell,0}$ and $\mathcal{A}_{r,0}$ by including one more operator $h_L$ with $\mathcal{A}_{\ell,0}$ and $h_R$ similarly for the right algebra. This extended (crossed product) algebra acts on an extended Hilbert space $\mathcal{H}= \mathcal{H}_{0}\otimes L^2(\mathbb{R})$ where the extra degree of freedom that has been introduced is the time-shift (the sum of the times in the left and the right exteriors). The extended crossed product right algebra is denoted $\mathcal{A}_{r}= \mathcal{A}_{r,0} \rtimes \mathbb{R}_{h}$ and similarly for the left algebra. Here,
  \begin{eqnarray}\label{II2}
 h_L= H_L - M_0  && h_R= H_R - M_0
 \end{eqnarray}
 $M_0$ is the ADM mass of some reference black hole.
 CPW work in a micro-canonical ensemble i.e. an energy eigenstate centered around some energy $M_0$ (mass of the reference black hole)\footnote{CPW explicitly do the microcanonical ensemble construction in the boundary CFT using a thermofield double state.}. The algebra of observables for the right exterior region is studied in a semi-classical limit i.e. $G\rightarrow 0$. In this limit, the ADM masses $H_R$ and $H_R$ diverge because the black hole mass $M_0$ (Schwarzschild radius divided by $2G$) diverges. So, CPW work with the non-divergent subtracted Hamiltonians $h_L$ and $h_R$.

 We can also write $\hat{h}=h_r - h_{\ell}$ where
 \begin{equation}\label{II3}
 h_r=\int_{S_1}d\Sigma^{\mu}V^{\nu} T_{\mu \nu}
 \end{equation}

 \begin{equation}\label{II4}
 h_{\ell}=-\int_{S_2}d\Sigma^{\mu}V^{\nu} T_{\mu \nu}
 \end{equation}
\begin{figure}[h]
  \centering
  \includegraphics[width=0.80\textwidth]{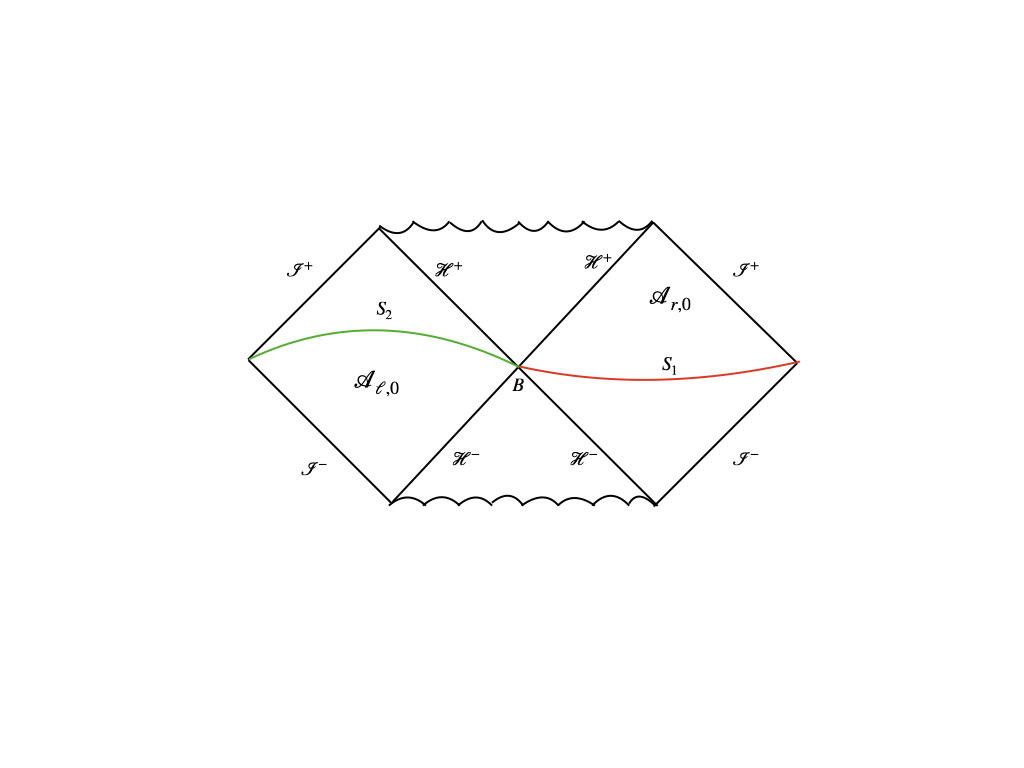}
  \caption{This figure depicts the split of Cauchy surface $S$ into union of the red Cauchy surface $S_1$ in the right exterior and the green Cauchy surface $S_2$  in the left exterior.}
  \label{fig:BH3}
\end{figure}
 where $S_1$ and $S_2$ are the right exterior and the left exterior part of the Cauchy surface $S$ as shown in Figure \ref{fig:BH3}.  As pointed out by CPLW, $ h_r$ and $h_\ell$ have divergent fluctuations. Thus, such a splitting is not true, strictly speaking \cite{EW3}, but in the extended algebra of Witten, the modular operator has a factorization into a product of operators in the left and right algebra.
   \begin{equation}\label{II5}
   \hat{h}= h_r - h_{\ell} =  h_R -h_L
   \end{equation}
    Further,
  \begin{equation}\label{II7}
  h_{R}= \hat{h}+h_{L}\equiv \frac{h_{\psi}}{\beta}+ x
  \end{equation}
where $h_{\Psi}= H_{mod}$ is the modular Hamiltonian for the Hartle-Hawking state $\ket{\Psi}= \ket{\Psi_{HH}}$.  $\mathcal{A}_{r}= \mathcal{A}_{r,0} \rtimes \mathbb{R}_{h}$ is the crossed product algebra of the algebra $\mathcal{A}_{r,0}$ by the modular group for the cyclic separating vector $\ket{\Psi}$. As discussed in \cite{EW3}, the algebra $\mathcal{A}_{r}$ is then a von Neumann algebra of  Type \RomanNumeralCaps {2}$_\infty$.  As explained in \cite{EW3}, unlike a Type \RomanNumeralCaps {3} algebra, a Type \RomanNumeralCaps {2} algebra has a notion of trace, density matrix and entropy.

An operator in $\mathcal{A}_{r}$ which we will denote as $\hat {a}$  has the form $\hat {a}= a e^{({is h_{\psi}/\beta}) }\otimes e^{(isx)}$, where $a \in \mathcal{A}_{r,0}$ . The states on which this operator acts has the form $\ket {\hat{\Psi}}=\ket{\Psi}\otimes g(x) \in \mathcal{H}$ where $\ket{\Psi} \in \mathcal{H}_0$ and $g(x)\in L^2(\mathbb{R})$. The most generic operator in $\mathcal{A}_r$ can be written as
\begin{equation}\label{II8}
\hat{a}=\int_{-\infty}^{\infty}a(s)e^{is(x+\hat{h})} ds
\end{equation}
 where $a(s)\in \mathcal{A}_{r,0}$. Similarly, we can write the most general state as
\begin{equation}\label{II9}
\ket{\hat{\Psi}}= \int dx f(x)\ket{\Psi} \ket{x}
\end{equation}
 As we have already  mentioned, a Type \RomanNumeralCaps {2} algebra has a trace, which is a positive linear functional on operators in the algebra satisfying $tr[\hat{a}\hat{b}]=tr[\hat{b}\hat{a}]$. The trace of an operator $\hat{a}\in \mathcal{A}_r$ can be denoted as \cite{VGE},
\begin{equation}\label{II10}
tr [\hat{a}]= \int_{-\infty}^{\infty} dx e^{\beta x}\bra{\hat{\Psi}}\hat{a}\ket{\hat{\Psi}}
\end{equation}
where the expectation value $\bra{\hat{\Psi}}\hat{a}\ket{\hat{\Psi}}$ is in general some non-trivial function of $x$. This trace is not the same as a standard trace on Hilbert space $\mathcal{H}$  but one can think of it more as a renormalized trace. Also, $tr$ is defined up to a scaling, which follows from the fact that the algebra $\mathcal{A}_r$ has a one-parameter family of outer automorphisms $x\rightarrow x+c$, which just rescales $tr$ by $e^c$. This rescaling has the physical interpretation of shifting the renormalization constant $M_0$ to $M_0 -c$. Now, using the trace, we can define the density matrix for the algebra $\mathcal{A}_r$. For any given state $\ket{\hat{\Phi}}\in \mathcal{H}$, the density matrix $\rho_{\hat{\Phi}} \in \mathcal{A}_{r}$ is defined by
\begin{equation}\label{II11}
tr[\rho_{\hat{\Phi}} \hat{a}]= \bra{\hat{\Phi}} \hat{a}\ket{\hat{\Phi}} \hspace{5mm} \forall \in \mathcal{A}_r
\end{equation}
  Moreover $\rho_{\hat{\Phi}}$ exists and is unique, which follows from the fact that the trace is non-degenerate. If we are able to define a density matrix, then we can also define the entropy of the algebra, by
  \begin{equation}\label{II12}
  S(\hat{\Phi})_{\mathbb{A}_r}= -tr[\rho_{\hat{\Phi}} \log{\rho_{\hat{\Phi}}}]= -\bra {\hat{\Phi}} \log {\rho_{\hat{\Phi}}} \ket{\hat{\Phi}}
\end{equation}
As emphasized by CPW,  $S(\hat{\Phi})$ should not be thought of as the entanglement entropy of $\mathcal{A}_r$ but is a renormalized entropy. Also, because of the ambiguity in the definition of trace (trace is defined up to a scaling), it is only entropy differences that are unambiguous, not the entropy itself. CPW now work with a semi-classical state, i.e the state with fluctuation in timeshift $p$, $\Delta p \sim O(\varepsilon)$, where $\varepsilon$ is some parameter much smaller than unity. Then, with $x=h_{L}$, $\Delta x \sim O(\frac{1}{\varepsilon})$. CPW consider the AdS-Schwarzschild black hole and write down the semi-classical state in the \emph{boundary} CFT.
However, it could have equally been defined in the bulk and we will assume that the formulae below correspond to the equivalent bulk statement.
The general form of such a state is,
\begin{equation}\label{II13}
\ket{\hat{\Phi}}= \int_{-\infty}^{\infty} \varepsilon^{\frac{1}{2}} g(\varepsilon x )\ket{\Phi}\ket{x} \hspace{3mm} \textrm{where} \hspace{5mm} \ket{\Phi} \in \mathcal{H},\hspace{2mm} g(x)\in L^2(\mathbb{R})
\end{equation}
It is shown by CPW that the density matrix $\rho_{\hat{\Phi}}$ for the state $\ket{\hat{\Phi}}$ is approximately \footnote{The expression is only valid up to corrections suppressed by $O(\varepsilon)$ terms.} given by
\begin{equation}\label{II14}
\rho_{\hat{\Phi}} \approx \varepsilon \bar{g}(\varepsilon h_{R})e^{-\beta x}\Delta_{\Phi|\Psi}g(\varepsilon h_{R})
\end{equation}
 where $\Delta_{\Phi|\Psi}= e^{-h_{\Psi|\Phi}}$ is a relative modular operator and $h_{\Psi|\Phi}$ is the relative modular Hamiltonian \footnote{$\Psi$ is the Hartle Hawking state and $\Phi$ is any arbitrary state of quantum fields in the black hole spacetime.}. The relative modular Hamiltonian $h_{\Phi|\Psi}$  is defined such that  $h_{\Psi|\Psi}= h_{\Psi}$. As we have already mentioned, the Type  \RomanNumeralCaps {2} algebra modular operator factorizes, i.e $\Delta_{\hat{\Phi}|\hat{\Psi}}= \rho_{\hat{\Psi}} \rho'^{-1}_{\hat{\Phi}}$ (where prime denotes the element of the commutant of the algebra $\mathcal{A}_{r}$)\cite{VRGE}. Putting (\ref{II14}) in (\ref{II12}) yields
 \begin{equation}\label{II15}
 S(\hat{\Phi})_{\mathcal{A}_{r}}=\bra{\hat{\Phi}}\beta h_{R}\ket{\hat{\Phi}}-\bra{\hat{\Phi}}h_{\Psi|\Phi}\ket{\hat{\Phi}}-\bra{\hat{\Phi}} \log{(\varepsilon |g(\varepsilon h_R)|^2)}\ket{\hat{\Phi}} + O(\varepsilon)
 \end{equation}
By definition, the second term in the above equation is the relative entropy,
\begin{equation}\label{II16}
S_{rel}(\Phi||\Psi)= -\bra{\hat{\Phi}}h_{\Psi|\Phi}\ket{\hat{\Phi}}
\end{equation}

\subsection{Generalization To Higher Curvature Gravity}
We now generalize the construction of the previous sub-section to an arbitrary diffeomorphism invariant theory of gravity. We note that some of the constructions in the previous sub-section such as the semi-classical state were done by CPW for the AdS-Schwarzschild black hole in the \emph{boundary} CFT. But we can analogously define such a semi-classical state in the bulk using the same construction. In fact, only in the sections on monotonicity of generalized entropy in CPW, are results in the boundary theory crucially used. Therefore, except while discussing monotonicity of the generalized entropy, we can confine our analysis to the bulk, and we can even work with an asymptotically flat black hole, as discussed by CPLW.
So, in our case, $M$ is the $(3+1)$ dimensional, asymptotically flat, maximally extended static (therefore stationary) black hole solution in an arbitrary diffeomorphism invariant theory of gravity. Therefore, its horizon is a Killing horizon. Now, the expectation value of the energy-momentum tensor of the quantum fields $\bra{{\Phi}}T_{\mu \nu} \ket{{\Phi}}$ is covariantly conserved as a consequence of invariance of the action under background diffeomorphisms (see appendix (\ref{a})). The equation (\ref{II1}) will thus define a conserved quantity even in the arbitrary theory of gravity if $V^{\mu}$ is a Killing vector.
Let us define the 1-form $J= T_{\mu \nu}V^{\nu}dx^\mu$ where $V$ is the timelike Killing vector of $M$. Then, divergence of $J_{\mu}$ is zero, i.e., $J^{\mu}$ is covariantly conserved. This implies $d* J=0$, where $*$ is the Hodge dual. This further implies $*J = dQ$ for some 2 form $Q$ \cite{compere}. Also, notice that the integral of $*J$ over the 3 dimensional Cauchy surface is  $\hat{h}$. Since $*J = dQ$, this reduces to an integral over the codimension 2 surface which is the boundary of the Cauchy slice. Therefore we can write $\hat{h}= H_R-H_L$ where $H_R$ and $H_L$ are codimension 2 integrals at right and left spatial infinity respectively. We note that the canonical energy $\mathcal{E}$ in the covariant phase space formalism is given by (\ref{II1}) apart from a surface term ambiguity (see appendix of IW \cite{VR}).
\begin{equation}\label{III}
\mathcal{E}= \int_{\Sigma} J + \text{Surface term}= \int_{\Sigma}d\Sigma^{\mu}T_{\mu \nu}V^{\nu} +\text{Surface term}
\end{equation}
 So we can think of $H_R$ and $H_L$ as being the right and left canonical energy, respectively, apart from ambiguities in the canonical energy \footnote{As discussed in the Appendix of IW \cite{VR}, the ambiguity in the canonical energy is the sum of two surface terms, one of which vanishes for common matter theories in a background spacetime. There is an ambiguity due to the second surface term which is a function of the background metric, Killing vector, matter fields and their derivatives. In what follows, we ignore this ambiguity.}. We will call them right and left Hamiltonian. Although the results of IW are only for classical gravity and matter, and we are interested in quantum fields, the IW equations can be understood as expectation values in semiclassical gravity as discussed in appendix (\ref{a}). Now, $\hat{h}$ is the modular Hamiltonian corresponding to the Hartle-Hawking state as before. This follows from the analysis of Sewell \cite{GS} for any metric of the following form:
\begin{equation}
ds^2 = A(t^2 - w^2, y)(- dt^2 + dw^2) + B(t^2 - w^2,y) d\sigma^2(y).
\end{equation}
The Schwarzschild spacetime in Kruskal coordinates is of this form. We will assume that our static black hole solution also has this form (i.e., we assume the existence of Kruskal-like coordinates).

Now we can proceed by defining $h_L$ and $h_R$ as defined in (\ref{II2}).  Following the argument in the previous section that including gravity changes the algebra to Type \RomanNumeralCaps {2}, we can split  $\hat{h}$ as done in (\ref{II3}) and (\ref{II4}). Further, we can straightforwardly obtain the equation in (\ref{II5}) and (\ref{II7}). The only difference is now $h_R$ in (\ref{II7}) is the renormalized Hamiltonian in the higher curvature theory which generates the time translation on the boundary of the right exterior region. Afterwards, the construction of the crossed -product algebra (extended algebra) and other constructions like defining the trace and entropy will analogously go through as done in the previous section. We will work with a semi-classical state as defined in (\ref{II13}). Therefore we can define the entropy of the algebra in the right exterior region by the same formula (\ref{II15}) i.e.
 \begin{equation}\label{III1}
 S(\hat{\Phi})_{\mathcal{A}_{r}}=\bra{\hat{\Phi}}\beta h_{R}\ket{\hat{\Phi}}-\bra{\hat{\Phi}}h_{\Psi|\Phi}\ket{\hat{\Phi}}-\bra{\hat{\Phi}} \log{(\varepsilon |g(\varepsilon h_R)|^2)}\ket{\hat{\Phi}} + O(\varepsilon)
 \end{equation}
where $S_{rel}(\Phi||\Psi)= -\bra{\hat{\Phi}}h_{\Psi|\Phi}\ket{\hat{\Phi}}$ is relative entropy as defined earlier. So the form of the equation (\ref{II15}) remains intact ---the only change is that $h_R$ is the renormalized canonical energy in higher curvature theory and $h_{\Phi|\Psi}$ is the relative modular Hamiltonian in that particular theory.
\\
In our case of interest, the black hole settles down to a stationary state at very late times. This is plausible since at late times all the flux of matter would either have crossed the horizon or would have escaped through future null infinity. So, at late times we will not be able to distinguish between $\ket{\Psi}$ and $\ket{\Phi}$. We get
\begin{equation}\label{II17}
S_{bulk}(\infty)_{\Phi}=S_{bulk}(\infty)_{\Psi}=S_{bulk}(b)_{\Psi}
\end{equation}
Now let us analyze $h_r$, using (\ref{II3}) and the fact that the deformation of Cauchy surfaces $S$ does not affect the conserved quantity $\hat{h}$. We deform $S_1$ such that $S'_1= \mathcal{H}^+ \cup \mathcal{I}^+$, where $\mathcal{H}^+$ is the future horizon and $\mathcal{I}^+$ is future null infinity \footnote{In the case of an asymptotically AdS black hole, the deformed Cauchy surface is just $\mathcal{H}^+$}. Therefore
\begin{equation}\label{II18}
\beta h_r^{\mathcal{H}^+}=\beta (h_r -h_r^{\mathcal{I^+}})= \int_{0}^{\infty} dv \int_{\mathcal{H}^{+}} d^{D-2}x \sqrt{h} v T_{vv}
\end{equation}
where $\beta h_r^{\mathcal{I^+}}$ is the time translation generator at future null infinity and $\beta h_r^{\mathcal{H}^+}$ is the boost generator on the horizon. The second equality in the above equation can be obtained using (\ref{II3})  and the fact that $h_r^{\mathcal{I^+}}$ is just the integral of the stress tensor supported at future null infinity. Let us define a one-sided modular operator (boost operator) at arbitrary cut $v=v_*$ ( which is the $D-2$ dimensional transverse surface) at the horizon. It is well known that the density matrix $(\rho_r)_{HH}$ of the Hartle Hawking state in the region $r$ is thermal with respect to \cite{JE,AW}
\begin{equation}\label{II18a}
K_r(v_*)=  \int_{v_*}^{\infty} dv \int_{\mathcal{H}^{+}} d^{D-2}x \sqrt{h} (v-v_*) T_{vv}+ K_r^{\mathcal{I^+}}
\end{equation}
where, $K_r^{\mathcal{I^+}}=\beta h_r^{\mathcal{I^+}}$ is the modular energy at $\mathcal{I^+}$, which accounts for energy which goes to $\mathcal{I^+}$ without crossing the horizon. Also notice that $K_r^{\mathcal{I^+}}$ is independent of $v_*$.  When $v_*=0$, then the first term in the above equation will become $\beta h_r^{\mathcal{H}^+}$ as defined in (\ref{II18}), and therefore
\begin{equation}\label{II18b}
K_r(b)=\beta h_r= \beta h_r^{\mathcal{H}^+} +\beta h_r^{\mathcal{I^+}}
\end{equation}
where $b$ is the bifurcation surface $v_*=0$. We can also define
\begin{equation}\label{II18c}
K_r(\infty)=\lim_{v_* \rightarrow \infty}K_r(v_*)
\end{equation}
Using the result from the previous section and equation (\ref{II18}), we get
\begin{equation}\label{II19}
\bra{\Phi }\beta h_r^{\mathcal{H^+}}\ket{\Phi}= \Delta \delta S
\end{equation}
If a density matrix were to exist for the algebra $\mathcal{A}_{r,0}$, then using the definition of modular Hamiltonian for state $\ket{\Psi}$ and the fact that $\Delta_{\Psi}= \rho_{\Psi}\rho'^{-1}_{\Psi} $, we will be able to write
\begin{equation}\label{II20}
\log{\rho_{\Psi}}=- K_{r}(b)+C
\end{equation}
where $C$ is some constant. The density matrix of the Hartle Hawking  state in region $r$ can be written as $\Psi(\mathcal{H^+}\cup \mathcal{I^+})= (\rho_r)_{HH}\otimes \sigma$, which corresponds to the ground state \footnote{Hartle Hawking state is a ground state with respect to the time $v$.} $\rho_{HH}$ at $\mathcal{H^+}$, product taken with some arbitrary density matrix defining a faithful state at $\mathcal{I^+}$ \cite{AW}. Therefore,
\begin{equation}\label{II20a}
\bra{\Psi}\log{\rho_{\Psi}}\ket{\Psi}= -\bra{\Psi}K_r^{\mathcal{I^+}}\ket{\Psi}+C
\end{equation}
Here we use the fact that $\bra{\Psi} h_r^{\mathcal{H^+}}\ket{\Psi}=0$ since it is the Hartle Hawking state. Further,  since $S_{bulk}(b)_{\Psi}=- \bra{\Psi}\log{\rho_{\Psi}}\ket{\Psi}$, we get $S_{bulk}(b)_{\Psi}= \bra{\Psi}K_r^{\mathcal{I^+}}\ket{\Psi}-C$.

As mentioned before, it is not strictly true of the algebra $\mathcal{A}_{r,0}$ that the modular operator factorizes, but by extending the algebra to $\mathcal{A}_{r}$, it is true that the modular operator factorizes as $\hat \Delta_{\Psi}= \rho_{\hat \Psi}\rho'^{-1}_{\hat \Psi} $ (in the notation of\cite{EW3}). We will ignore this detail just for illustrative purposes following \cite{VGE} .
\begin{equation}\label{II20aa}
S_{gen}(\infty)-S_{gen}(b) = S(\infty) - S(b) + S_{bulk}(b)_{\Psi} - S_{bulk}(b)_{\Phi}.
\end{equation}
It can be written in terms of the one-sided modular operator,
\begin{equation}\label{II20i}
S_{gen}(\infty)-S_{gen}(b) = - \bra{\Phi} (K_{r}(\infty)-K_r(b)) \ket{\Phi} + S_{bulk}(b)_{\Psi} - S_{bulk}(b)_{\Phi}.
\end{equation}
Putting $S_{bulk}(b)_{\Psi}$ and $\bra{\Phi}K_r(b)\ket{\Phi}$ using the equation (\ref{II20}) , we get
\begin{equation}\label{II20ia}
S_{gen}(\infty)-S_{gen}(b) = -\bra{\Phi} \log{\rho_{\Psi}} \ket{\Phi}+ C -\bra{\Phi} K_{r}(\infty)\ket{\Phi} + \bra{\Psi} K_r^{\mathcal{I^+}} \ket{\Psi}-C -S_{bulk}(b)_{\Phi}
\end{equation}
Now, we use the fact that at late times, every state is indistinguishable from $\ket{\Psi}$ and $K_r(\infty)=K_r^{\mathcal{I^+}}$. Further, $K_r^{\mathcal{I^+}}$  is independent of the cut. The expectation value $ K_r^{\mathcal{I^+}}$ in state $\ket{\Phi}$ will be equal to its expectation value in state $\ket{\Psi}$. Therefore, we get
\begin{equation}\label{II20ii}
S_{gen}(\infty)-S_{gen}(b) = - \bra{\Phi}\log{\rho_{\Psi}} \ket{\Phi } - S_{bulk}(b)_{\Phi}
\end{equation}
therefore we got,
\begin{equation}\label{II20b}
S_{gen}(\infty)-S_{gen}(b)=S_{rel}(\Phi||\Psi)
\end{equation}
As we see, the difference of generalized entropies in (\ref{II20b}) is manifestly finite and non-negative.
For Einstein gravity, the above expression has been already obtained by Wall in \cite{AW}. The result (\ref{II20b}) is in an arbitrary theory of gravity --- the difference between generalized entropy at late times and generalized entropy at the bifurcation surface is relative entropy of the state of the black hole with respect to the Hartle Hawking state. We now need to show, as in \cite{VGE}, that the generalized entropy at the bifurcation surface is the entropy of the algebra. We thus need to show
 \begin{equation}\label{II21}
S_{gen}(\infty)= \bra{\hat{\Phi}}\beta h_{R}\ket{\hat{\Phi}}-\bra{\hat{\Phi}} \log{(\varepsilon |g(\varepsilon h_R)|^2)}\ket{\hat{\Phi}}+ \textrm{Const}
 \end{equation}
 Since both terms in the above equation are only functions of $h_{R}$, and since we have interpreted $h_{R}$ as the renormalized canonical energy, the above terms are some distributions of energy in the semi-classical state $\ket{\hat{\Phi}}$. Also, these terms are independent of the state $\ket{\Phi}$. To see that, choose $a(s)$ such that
 \begin{equation}\label{II21a}
 a(s)=\int e^{-ih'_{R}s} f(h'_{R}) dh'_{R}
 \end{equation}
where $f(h_{R})$ is some chosen function. Putting the equation (\ref{II21a}) in the equation (\ref{II8}) and using the fact that $h_R=\hat{h_{\Psi}}+x$, will yield $\hat{a}= f(h_R)$. Now let us compute the expectation value $\bra{\hat{\Phi}}\beta \hat{a}\ket{\hat{\Phi}}$ for (\ref{II21a}) with the semi-classical state $\ket{\hat{\Phi}}$ defined in (\ref{II13}). Using the results (3.25) and (3.26) in CPW \cite{VGE},
  \begin{equation}\label{II21b}
  \bra{\hat{\Phi}}\hat{a}\ket{\hat{\Phi}}= \int_{-\infty}^{\infty}dx \int_{-\infty}^{\infty}ds |\varepsilon g(\varepsilon x)|^2 e^{isx} \bra{\Psi}\Delta_{\Phi|\Psi}a(s)\ket{\Psi}
  \end{equation}
 Now put (\ref{II21a}) in (\ref{II21b}). Using the fact that $h_R=\hat{h_{\Psi}}+x$ and $\hat{h_{\Psi}}\ket{\Psi}=0$, we get
 \begin{equation}\label{II21c}
 \bra{\hat{\Phi}} f(h_R) \ket{\hat{\Phi}}= \int_{-\infty}^{\infty}dx \int_{-\infty}^{\infty}ds \int_{-\infty}^{\infty} dy |\varepsilon g(\epsilon x)|^2 e^{is(x-y)} f(y) \bra{\Psi}\Delta_{\Phi|\Psi}\ket{\Psi}
\end{equation}
By definition, $\bra{\Psi}\Delta_{\Phi|\Psi}\ket{\Psi}=1$. Therefore the above equation is independent of $\ket{\Phi}$, it will only depend on $f(h_R)$ and $g(\varepsilon x)$. Therefore, both the terms on the right-hand side of (\ref{II21}) will give the same result either when we compute them in the state $\ket{\Phi}$ or in the Hartle Hawking state $\ket{\Psi}$ at late times.
Since both terms can be determined from the late-time behavior of the black holes, the relation (\ref{II21}) is plausible.
This is because at late times, all the fields have either fallen across the horizon or to infinity.
 \\
 Using equation (\ref{II2}), we can write $\bra{\hat{\Psi}}\beta h_{R}\ket{\hat{\Psi}}= \beta(\Delta\mathcal{E}) $, where $\Delta\mathcal{E}$ is the energy difference between the black hole we are studying and the reference black hole. Since both the black holes are taken in an equilibrium state, we can apply the first law of black hole mechanics for two equilibrium configurations in phase space which yields
 \begin{equation}\label{II22}
 \bra{\hat{\Psi}}\beta h_{R}\ket{\hat{\Psi}}= \Delta S
 \end{equation}
  where $\Delta S$ is the difference of entropy of the equilibrium black hole state we get at late times to the reference black hole. Therefore, the first term in (\ref{II21}) describes the change in  $S_{gen}(\infty)$ due to a change in black hole entropy. At very late times, all the matter would have either crossed the horizon or would have escaped to null infinity. The second term should be thought of as the contribution of entropy of fluctuations in black hole entropy \cite{VGE}. Finally, we add and subtract the entanglement entropy of the quantum fields in the Hartle-Hawking state and lump one of the pieces in the constant in (\ref{II21}) using (\ref{II20a}). This is because at late times, all the fields have either fallen across the horizon or to infinity.

  Combining everything, we get
  \begin{equation}\label{II23}
  S(\hat{\Phi})_{\mathcal{A}_r}= S_{gen}(b)+ \textrm{Const} .
\end{equation}
  The equation (\ref{II23}) tells us that $S_{gen}(b)$ for black holes at the bifurcation surface in the arbitrary theory of gravity can be thought of as the entropy of the algebra $\mathcal{A}_r$ modulo a constant. Notice, we have shown that the generalized entropy at the bifurcation surface is equal to the entropy of the algebra up to a constant, but we are not making any statement about entropy at an arbitrary cut of the horizon. In algebraic QFT, relative entropy is positive. This implies $S_{gen}(\infty)-S_{gen}(b) \geq 0$. Can we go beyond this result and prove that the generalized entropy is monotonic? The entropy of the algebra is monotonic under trace-preserving inclusions \cite{LW}. To obtain a GSL (monotonicity of the generalized entropy), CPW consider an AdS Schwarzschild black hole with a holographic dual CFT. Then, they have the following clever argument: In the dual CFT, they consider operator algebras at two different times (early and late times), $\mathcal{A}_{R,0}$ and $\mathcal{B}_{R,0}$ respectively, separated by a timescale much larger than the thermal time scale $\beta$. The correlation functions of operators at these different times factorize into a product of correlators of early and late times. Thus, the algebra generated by both early and late time operators is $\mathcal{C}_{R,0} = \mathcal{A}_{R,0} \otimes \mathcal{B}_{R,0}$. The Hilbert space factorizes similarly. These algebras are extended similar to what was done before, to obtain a Type \RomanNumeralCaps {2} algebra which has an associated entropy. Now, consider three different situations: first, the quantum fields at both early and late times are in an arbitrary state, a second situation where the fields at early times have fallen into the horizon, so that the state of these fields is the vacuum times any state of the late time fields, and finally, a situation where both sets of fields have fallen into the horizon and the state of the fields in the exterior is the vacuum. CPW argue that the generalized entropies at these three horizon cuts is the generalized entropy for the extended algebra $\mathcal{C}_{R}$ for these three different states at a hypothetical  \emph{bifurcation surface} in the limit of large time gap. From the property of the monotonicity of the entropy of the algebra under trace preserving inclusions, it follows that the generalized entropy is increasing in going from the first to the third situation above. This argument is then a version of the GSL. We can use these results to prove this version of the GSL in an arbitrary diffeomorphism invariant theory of gravity if we start with an asymptotically AdS black hole which has a holographic dual. We can repeat all the steps in this section for such a black hole. The only thing we need is for the first law as in the paper of Iyer and Wald \cite{VR} to be true in this situation. Although the statement of the First law is only for asymptotically flat black holes, the same will be true for an asymptotically AdS black hole provided the integrals involved in the presymplectic form and the canonical energy are finite after assuming appropriate fall-offs for the fields. In this case, the computation of CPW generalizes to these black holes in a higher curvature theory of gravity and a version of increase of generalized entropy (comparing the entropy at early and late times) is true. The ambiguity in the Wald entropy which we fixed in the bifurcation surface in a previous section was by terms quadratic in the half-order perturbation. This is not affected by the different states of the quantum fields in the argument of CPW, so their argument goes over to our case. Can we show a stronger monotonicity result $\frac{dS_{gen}}{dv} \geq 0$? This is what Wall \cite{AW} has done for Einstein gravity, using an expression for the entropy at any cut of the horizon. Due to JKM ambiguities in the Wald entropy, this expression will probably need one to specify the particular theory of gravity.

\section{Discussion}
In the context of Einstein gravity, it was shown by CPW \cite{VGE} that for the system of quantum fields in a perturbed Schwarzschild black hole spacetime in the $G \to 0$ limit with infalling quantum matter across the horizon, the generalized entropy at the bifurcation surface was equal to the entropy of the von Neumann algebra of operators in the black hole exterior. This was achieved by enlarging the operator algebra by including the ADM Hamiltonian, and by enlarging the Hilbert space. This had the effect of changing the algebra of operators to a Type \RomanNumeralCaps {2}$_\infty$ von Neumann algebra, to which we can associate a notion of trace and entropy. Furthermore, CPW showed that the difference of the generalized entropy of an arbitrary cut of the horizon, in the limit when the cut $v \to \infty$ and the generalized entropy at the bifurcation surface was equal to the relative entropy, and therefore nonnegative. For this, CPW worked in semiclassical gravity and up to quadratic order in perturbations. They then obtained a monotonicity result (GSL) for the generalized entropy from the monotonicity of relative entropy under trace preserving inclusions.
In this paper, we consider quantum fields in a slightly perturbed static black hole in an arbitrary diffeomorphism invariant theory of gravity in the $G \to 0$ limit. In this set-up, generalized entropy is the sum of Wald entropy and entanglement entropy of quantum fields in the black hole exterior.  We consider the difference in Wald entropy at infinity and at the bifurcation surface up to quadratic order in the perturbations and obtain (\ref{I45}). Wald entropy has ambiguities on non-stationary geometries. We fix the ambiguity in order to get (\ref{I46}) which matches the result for Einstein gravity in the paper of CPW and we obtain a simplified result for the difference of entropies which enables us to employ the CPW construction. We then consider the difference in \emph{generalized} entropies and show that this difference equals the relative entropy of the state of the quantum fields and the Hartle-Hawking state --- it is thus non-negative. We next consider the von Neumann algebra of the quantum fields in the black hole exterior, extended to include the Hamiltonian, and an enlarged Hilbert space as in CPW. Evaluated on the semiclassical states defined by CPW, we show that the entropy of the von Neumann algebra equals the generalized entropy of the bifurcation surface. Finally, we see that the derivation of the increase of generalized entropy by CPW in Einstein gravity goes through for black holes in an arbitrary gravity theory, provided the black hole is asymptotically AdS, which has a holographic dual.

Going forward, it would be interesting to not merely restrict to semiclassical states and see what the entropy of the algebra corresponding to more general states is. Recently, there has also been an interesting proposal to study the algebra of operators in general subregions in Einstein gravity with matter in the $G \to 0$ limit, and it has been shown that the entropy of this algebra equals the generalized entropy of this subregion up to a constant \cite{sorce} (see also the latest papers \cite{jefferson} and \cite{leigh}). It would be interesting to see how these results generalize to an arbitrary diffeomorphism invariant theory of gravity.
\section{Acknowledgements} We would like to thank Sanved Kolekar for useful comments. MA also acknowledges the Council of Scientific and Industrial Research (CSIR), Government of India for financial assistance.
\section{Appendix A}\label{a}
\subsection{Quantum Canonical energy in covariant phase space formalism}
Consider quantum fields in the stationary black hole background, where the quantum fields include gravitons. Using diffeomorphism invariance of the Lagrangian (which is an $n$-form) with respect to the background metric, we can write
\begin{equation}\label{IIIa1}
 \frac{\delta L}{\delta \phi}\mathcal{L}_{\xi}\phi= -d\tilde J- \frac{\delta L}{\delta g_{ab}}\mathcal{L}_{\xi}g_{ab} .
\end{equation}
$\phi$ here corresponds to all quantum fields and $g_{ab}$ is the background metric. $\xi$ is a vector field generating the diffeomorphism and $\mathcal{L}_{\xi}$ denotes the Lie derivative with respect to $\xi$.
Here, $\tilde J$ is the Noether current $(n-1)$ form in the covariant phase space formalism given by equation (49) in IW \cite{VR}. This is defined in terms of the symplectic potential form $\Theta(\delta \phi, \phi)$ as
\begin{equation}\label{IIIc}
\tilde J = \Theta(\phi, \mathcal{L}_{\xi} \phi ) - \xi \cdot L .
\end{equation}

As we have a Lagrangian for quantum fields to begin with, we look at the expectation value of the relation (\ref{IIIa1}) in an arbitrary state. Since the terms in the equations are products of operators at the same point, the expectation values need to be regularized using some procedure. Using a prescription involving point-splitting and background subtraction, we regularize these expectation values.

We want to compute the expectation value of (\ref{IIIa1}) in some state $\ket{\Phi}$. Consider the point-split expression
 \begin{equation}\label{IIIa2}
\lim_{y\rightarrow x} \bra{\Phi}\frac{\delta L}{\delta \phi}(x) \mathcal{L}_{\xi}\phi (y) \ket{\Phi} =\lim_{y\rightarrow x}\bra{\Phi}\Big(- d\tilde J(x,y)- \frac{\delta L}{\delta g_{ab}}(x)\mathcal{L}_{\xi}g_{ab}(y)\Big) \ket{\Phi}
 \end{equation}
From the Schwinger Dyson equation for expectation value in an arbitrary state, we can compute the left hand side, and in the limit of coincident points, it is just a state independent divergent term. Therefore if we consider the difference of the quantity  $\Big(- d\tilde J(x,y)- \frac{\delta L}{\delta g}(x)\mathcal{L}_{\xi}g(y)\Big)$ in any two states $\ket{\Phi}$ and $\ket{\Psi}$, then the state independent divergent term will cancel out. Now we can take the coincidence limit  $y\rightarrow x$.
\begin{equation}\label{IIIa3}
\bra{\Phi}\Big( -d\tilde J(x)- \frac{\delta L}{\delta g_{ab}}(x)\mathcal{L}_{\xi}g_{ab}(x)\Big) \ket{\Phi}- \bra{\Psi}\Big( -d\tilde J(x)- \frac{\delta L}{\delta g_{ab}}(x)\mathcal{L}_{\xi}g_{ab}(x)\Big) \ket{\Psi}=0
\end{equation}
Now, (\ref{IIIa3}) can be written as
\begin{equation}\label{IIIa4}
d\bra{\Phi} (\tilde J +k.\epsilon)\ket{\Phi}-d\bra{\Psi} (\tilde J +k.\epsilon)\ket{\Psi}=\nabla_\mu (\bra{\Phi}  T^{\mu \nu}\ket{\Phi} )\xi_{\nu}-\nabla_\mu (\bra{\Psi}  T^{\mu \nu}\ket{\Psi}) \xi_{\nu}.
\end{equation}
Here, we use the fact that $\frac{\delta L}{\delta g_{\mu \nu}}= T^{\mu \nu} \epsilon$, where $\epsilon$ is the volume form and $k^\mu= T^{\mu \nu} \xi_{\nu}$. Notice that the left hand side in the above equation is a total derivative for any $\xi$ while the right hand side is not. The only way that can happen is if $\nabla_\mu (\bra{\Phi}  T^{\mu \nu}\ket{\Phi}) \xi_{\nu}=\nabla_\mu (\bra{\Psi}  T^{\mu \nu}\ket{\Psi}) \xi_{\nu}$. Since this has to be true for any two states and any vector field $\xi^{\mu}$, it can only be if $ \nabla_\mu (\bra{\Phi}  T^{\mu \nu}\ket{\Phi}) $ vanishes for any state $\ket{\Phi}$ up to a local term independent of state. We may modify our prescription and get rid of this extra state independent term by doing background subtraction with respect to some standard state. We do the background subtraction (\ref{IIIa3}) to make sense of expectation values of equation (\ref{IIIa1}).  Thus the expectation value of the stress energy tensor in any state is conserved. It has already been extensively discussed in Wald \cite{W} how to define the expectation value of the stress energy tensor in such a way that it is conserved, so alternatively, we could use that result. By the same argument, $d\bra{\Phi} (\tilde J +k.\epsilon)\ket{\Phi}=0$. The rest of the analysis of the Appendix of IW \cite{VR} follows. Now we can choose $\xi^{\mu}$ to be the  Killing field of the static background spacetime. Following IW \cite{VR}, we will get
\begin{equation}\label{IIIa5}
\big<\mathcal{E}\big>_{\Phi}=\int_{\Sigma} d\Sigma^{\mu}\big<T_{\mu \nu}\big>_{\Phi}\xi^{\nu} + \text{Surface term}
\end{equation}
where $\mathcal{E}$ is the canonical energy in the covariant phase space formalism and it is independent of the choice of Cauchy slice.
 
\end{document}